\documentclass[aps,prd,eqsecnum,preprint,tightenlines,nofootinbib]{revtex4-1}
\usepackage{graphicx,latexsym}

\newcommand{\bea}{\begin{eqnarray}}
\newcommand{\eea}{\end{eqnarray}}
\newcommand{\be}{\begin{equation}}
\newcommand{\ee}{\end{equation}}

\newcommand{\Pminus}{{\cal P}^-}

\newcommand{\pp}{p^{\prime +}}
\newcommand{\Nmax}{N_{\rm max}}

\newcommand{\rvac}{|0\rangle}

\begin{document}

\title{Symmetry breaking in light-front $\phi^4$ theory%
\footnote{Based on a talk contributed to the
Lightcone 2016 workshop, Lisbon, Portugal, 
September 5-8, 2016.}
}

\author{J.R. Hiller}

\affiliation{Department of Physics and Astronomy\\
University of Minnesota-Duluth \\
Duluth, Minnesota 55812}

\date{\today}

\begin{abstract}
We consider the symmetric and broken phases of light-front
$\phi^4$ theory in two dimensions.  In both cases the
mass of the lowest state is computed and its dependence
on the coupling used to infer critical coupling values.
The structure of the eigenstate is examined to determine
whether it shows the signs of critical behavior, specifically
whether the one-body sector becomes improbable relative
to the higher Fock sectors.  In attempts to establish
this behavior, we consider both sector-independent and
sector-dependent constituent masses.
\end{abstract}

\maketitle

\section{Introduction} \label{sec:intro}

We apply a new computational method to light-front~\cite{LFreview1,LFreview2,LFreview3,LFreview4}
$\phi^4$ theory in two dimensions, in both the
symmetric and broken phases~\cite{RozowskyThorn,Kim,Varyetal1,Varyetal2,Varyetal3,Varyetal4}.  
The method is based on an expansion of the Fock-state wave functions
in a basis of multivariate symmetric polynomials~\cite{GenSymPolys1,GenSymPolys2}.
This allows fine tuning of the resolution, Fock sector by
Fock sector, and incorporates small-$x$ behavior that
captures an integrable singularity.  Both features represent
an improvement over the traditional discrete light-cone
quantization (DLCQ)~\cite{PauliBrodsky1,PauliBrodsky2} approach, where the resolution
is fixed across all Fock sectors and the integrable singularity
at zero momentum fraction is ignored.  The presentation here
extends earlier work in \cite{phi4sympolys}.

The general method and an application to
the broken phase are described here; the
symmetric phase is discussed specifically
by Chabysheva~\cite{ChabyshevaLC16}.
Section~\ref{sec:formulation} details the
structure of the $\phi^4$ eigenvalue problem
and our method of solution, including the
option of a sector-dependent mass.  Results
are presented in Sec.~\ref{sec:results},
followed by a brief summary in Sec.~\ref{sec:summary}

\section{Eigenvalue problem}  \label{sec:formulation}

The Lagrangian for $\phi^4$ theory is
${\cal L}=\frac12(\partial_\mu\phi)^2-\frac12\mu_0^2\phi^2-\frac{\lambda}{4!}\phi^4$.
From it, one obtains the light-front Hamiltonian density
${\cal H}^-=\pm\frac12 \mu^2 :\phi^2:+\frac{\lambda}{4!}:\phi^4:$
and the Hamiltonian $\Pminus=\int dx^- {\cal H}^-$
that defines the eigenvalue problem $\Pminus|\psi\rangle=\frac{M^2}{P^+}|\psi\rangle$.
Here, $P^+=E+p_z$ is the light-front momentum conjugate to the light-front
spatial coordinate $x^-\equiv t-z$, and $\Pminus$ is the operator that generates 
translations in light-front time $x^+\equiv t+z$.  The mass of the
eigenstate $|\psi\rangle$ is $M$.

The Hamiltonian is computed with the mode expansion
\be \label{eq:mode}
\phi(x^+,x^-)=\int \frac{dp^+}{\sqrt{4\pi p^+}}
   \left\{ a(p^+)e^{-ipx} + a^\dagger(p^+)e^{ipx}\right\}.
\ee
The creation operators $a^\dagger(p^+)$ satisfy the
commutation relation
$[a(p^+),a^\dagger(\pp)]=\delta(p^+-\pp)$.

However, before completing the construction of $\Pminus$, we 
consider an asymmetric form obtained by a shift in the field.
The generator for the shift is
$U=\exp\int dp^+ [f(p^+)a^\dagger(p^+)-f^*(p^+)a(p^+)]$.
This shifts the creation operator 
$Ua^\dagger(p^+) U^\dagger=a^\dagger(p^+)+f^*(p^+)$.
Following Harindranath and Vary~\cite{HariVary}, we choose $f$ 
to correspond to a zero mode
$f(p^+)=f^*(p^+)\equiv\sqrt{\pi p^+}\delta(p^+)\phi_s$.  The
field is then shifted by a constant, $U\phi U^\dagger=\phi+\phi_s$ ,
and, with $\phi_s=\pm\sqrt{6\mu^2/\lambda}$, we obtain
\be
U:{\cal H}^-\!: U^\dagger=-\frac32\frac{\mu^4}{\lambda}+\frac12(2\mu^2):\phi^2:
    +\frac{\lambda\phi_s}{3!}:\phi^3:+\frac{\lambda}{4!}:\phi^4:
\ee

Now substitution of the mode expansion yields
$\Pminus=\Pminus_{11}+\Pminus_{22}+\Pminus_{13}+\Pminus_{31}+\Pminus_{12}+\Pminus_{21}$,
with
\bea
\Pminus_{11}&=&\int dp \frac{2\mu^2}{p} a^\dagger(p)a(p),  \\
\Pminus_{22}&=&\frac{\lambda}{4}\int\frac{dp_1 dp_2}{4\pi\sqrt{p_1p_2}}
       \int\frac{dp'_1 dp'_2}{\sqrt{p'_1 p'_2}} 
       \delta(p_1 + p_2-p'_1-p'_2) a^\dagger(p_1) a^\dagger(p_2) a(p'_1) a(p'_2), \\
\Pminus_{13}&=&\frac{\lambda}{6}\int \frac{dp_1dp_2dp_3}
                              {4\pi \sqrt{p_1p_2p_3(p_1+p_2+p_3)}} 
     a^\dagger(p_1+p_2+p_3)a(p_1)a(p_2)a(p_3), \\
\Pminus_{12}&=& \mu\sqrt{\frac{3\lambda}{2}}
   \int \frac{dp_1^+ dp_2^+}{\sqrt{4\pi p_1^+ p_2^+(p_1^++p_2^+)}}
       a^\dagger(p_1^++p_2^+)a(p_1^+)a(p_2^+), \\
\Pminus_{31}&=&(\Pminus_{13})^\dagger,\;\;\Pminus_{21}=(\Pminus_{12})^\dagger.
\eea

The eigenstate of $\Pminus$, with eigenvalue $M^2/P^+$,
can be expressed as an expansion 
\be \label{eq:FSexpansion}
|\psi(P^+)\rangle=\sum_m P^{+\frac{m-1}{2}}\int\prod_i^m dy_i 
       \delta(1-\sum_i^m y_i)\psi_m(y_i)|y_iP^+;P^+,m\rangle
\ee
in terms of Fock states
$|y_iP^+;P^+,m\rangle=\frac{1}{\sqrt{m!}}\prod_{i=1}^m a^\dagger(y_iP^+)\rvac$
with normalization
$1=\sum_m \int\prod_i^m dy_i \delta(1-\sum_i^m y_i)|\psi_m(y_i)|^2$.
The eigenvalue problem is then reduced to a coupled system of equations
\bea
\lefteqn{\left\{\begin{array}{l} +\mu^2 \\ -\mu^2 \\ 2\mu^2\end{array}\right\}
  \sum_i^m \frac{1}{y_i }\psi_m(y_i)
+\frac{\lambda}{4\pi}\frac{m(m-1)}{4\sqrt{y_1y_2}}
        \int\frac{dx_1 \psi_m(x_1,y_1+y_2-x_1,y_3,\ldots,y_m)}{\sqrt{x_1 (y_1+y_2-x_1)}}} && \\
&& +\frac{\lambda}{4\pi}\frac{m\sqrt{(m+2)(m+1)}}{6}
       \int \frac{dx_1 dx_2 \psi_{m+2}(x_1,x_2,y_1-x_1-x_2,y_2,\ldots,y_m)}{\sqrt{y_1 x_1 x_2 (y_1-x_1-x_2)}}
       \nonumber \\
&&+\frac{\lambda}{4\pi}\frac{(m-2)\sqrt{m(m-1)}}{6}
      \frac{\psi_{m-2}(y_1+y_2+y_3,y_4,\ldots,y_m)}{\sqrt{y_1y_2y_3(y_1+y_2+y_3)}} \nonumber \\
&& +\mu\sqrt{\frac{3\lambda}{8\pi}}m\sqrt{m+1}
    \int \frac{dx_1 \psi_{m+1}(x_1,y_1-x_1,y_2,\ldots,y_m)}{\sqrt{x_1 (y_1-x_1) y_1}} \nonumber \\
&& +\mu\sqrt{\frac{3\lambda}{8\pi}}(m-1)\sqrt{m} 
  \frac{\psi_{m-1}(y_1+y_2,y_3,\ldots,y_m)}{\sqrt{y_1 y_2 (y_1+y_2)}}=M^2\psi_m(y_i), \nonumber
\eea
where the last two terms are kept only for bottom option of $2\mu^2$, which represents the
shifted, asymmetric form of the Hamiltonian.  In the symmetric phase, we consider only the
odd eigenstate where the number of constituents $m$ is an odd integer.

We solve this system by truncation in Fock space, to $m\leq\Nmax$, and by expansion 
of the Fock-state wave functions in terms of multivariate symmetric 
polynomials $P_{ni}^{(m)}(y_1,\ldots,y_m)$~\cite{GenSymPolys1,GenSymPolys2}
\be \label{eq:expansion}
\psi_m(y_1,\ldots,y_m)=\sqrt{y_1 y_2\cdots y_m}\sum_{ni} c_{ni}^{(m)} P_{ni}^{(m)}(y_1,\ldots,y_m).
\ee
This converts the system into a generalized matrix eigenvalue problem
\bea \label{eq:matrixequations}
\lefteqn{\sum_{n'i'}\left[\left\{\begin{array}{l} +1 \\ -1 \\ 2\end{array}\right\}
    T^{(m)}_{ni,n'i'}+ gV^{(m,m)}_{ni,n'i'}\right]c^{(m)}_{n'i'}
  + g\sum_{n'i'} V^{(m,m+2)}_{ni,n'i'} c^{(m+2)}_{n'i'}
   + g\sum_{n'i'} V^{(m,m-2)}_{ni,n'i'} c^{(m-2)}_{n'i'}} && \\
 &&  + \sqrt{g}\sum_{n'i'} V^{(m,m+1)}_{ni,n'i'} c^{(m+1)}_{n'i'}
   + \sqrt{g}\sum_{n'i'} V^{(m,m-1)}_{ni,n'i'} c^{(m-1)}_{n'i'} 
  =\frac{M^2}{\mu^2}\sum_{n'i'}B^{(m)}_{ni,n'i'}c_{n'i'}^{(m)}, \nonumber
\eea
with $g=\lambda/(4\pi\mu^2)$.  The matrices $T$, $V$ and $B$ are defined as
\be
T^{(m)}_{ni,n'i'}=m\int\left(\prod_j dy_j \right)\delta(1-\sum_j y_j)\left(\prod_{j=2}^m y_j\right) P_{ni}^{(m)}(y_j)P_{n'i'}^{(m)}(y_j),
\ee
\bea
V^{(m,m)}_{ni,n'i'}&=&\frac{1}{4}m(m-1)\int\left(\prod_j dy_j\right) \delta(1-\sum_j y_j) \\
  && \times  \int dx_1 dx_2 \delta(y_1+y_2-x_1-x_2)
      \left(\prod_{j=3}^m y_j\right) P_{ni}^{(m)}(y_j)P_{n'i'}^{(m)}(x_1,x_2,y_3,\ldots,y_m),
      \nonumber
\eea
\bea
\lefteqn{V^{(m,m+2)}_{ni,n'i'}=\frac{1}{6}m\sqrt{(m+2)(m+1)}\int\left(\prod_j dy_j\right) \delta(1-\sum_j y_j)}&& \\
  && \times 
    \int dx_1 dx_2 dx_3 \delta(y_1-x_1-x_2-x_3)
      \left(\prod_{j=2}^m y_j\right) P_{ni}^{(m)}(y_j)P_{n'i'}^{(m+2)}(x_1,x_2,x_3,y_2,\ldots,y_m),
      \nonumber
\eea
%
%

\bea
V^{(m,m+1)}_{ni,n'i'}&=&\sqrt{\frac32}m\sqrt{m+1}\int\left(\prod_j dy_j\right) \delta(1-\sum_j y_j) \\
  && \times 
    \int dx_1 dx_2 \delta(y_1-x_1-x_2)
      \left(\prod_{j=2}^m y_j\right) P_{ni}^{(m)}(y_j)P_{n'i'}^{(m+1)}(x_1,x_2,y_2,\ldots,y_m),
      \nonumber
\eea
%
%
and
\be
B^{(m)}_{ni,n'i'}=\int\left(\prod_j dy_j \right)\delta(1-\sum_j y_j)\left(\prod_j^m y_j\right) P_{ni}^{(m)}(y_j)P_{n'i'}^{(m)}(y_j),
\ee
with $V^{(m,m-2)}_{ni,n'i'}$ and $V^{(m,m-1)}_{ni,n'i'}$ obtained as the adjoints
of $V^{(m-2,m)}_{ni,n'i'}$ and $V^{(m-1,m)}_{ni,n'i'}$, respectively.

We convert the matrix problem to an ordinary eigenvalue problem by a singular value
decomposition (SVD)~\cite{SVD} of the overlap matrix $B^{(m)}$.  This is an implicit 
orthogonalization of the basis.  The SVD is $B^{(m)}=U^{(m)}D^{(m)}U^{(m)T}$, where the 
columns of $U^{(m)}$ are the eigenvectors of $B^{(m)}$ and 
$D^{(m)}$ is a diagonal matrix of the eigenvalues.
We keep in $U^{(m)}$ only those columns associated with eigenvalues of $B^{(m)}$
that are above some positive threshold,
in order to eliminate from the basis those combinations of polynomials that are nearly
linearly dependent~\cite{Wilson}.
We then define new vectors of coefficients $\vec c^{\,(m)\prime}=D^{1/2}U^T\vec c^{\,(m)}$
and new matrices, such as $T^{(m)\prime}=D^{-1/2}U^T T^{(m)} UD^{-1/2}$.  In terms of
these, the matrix eigenvalue problem is an ordinary one, which we then diagonalize by
standard means.

\section{Results} \label{sec:results}

\begin{figure}
\centering
\begin{tabular}{cc}
\includegraphics[width=7cm]{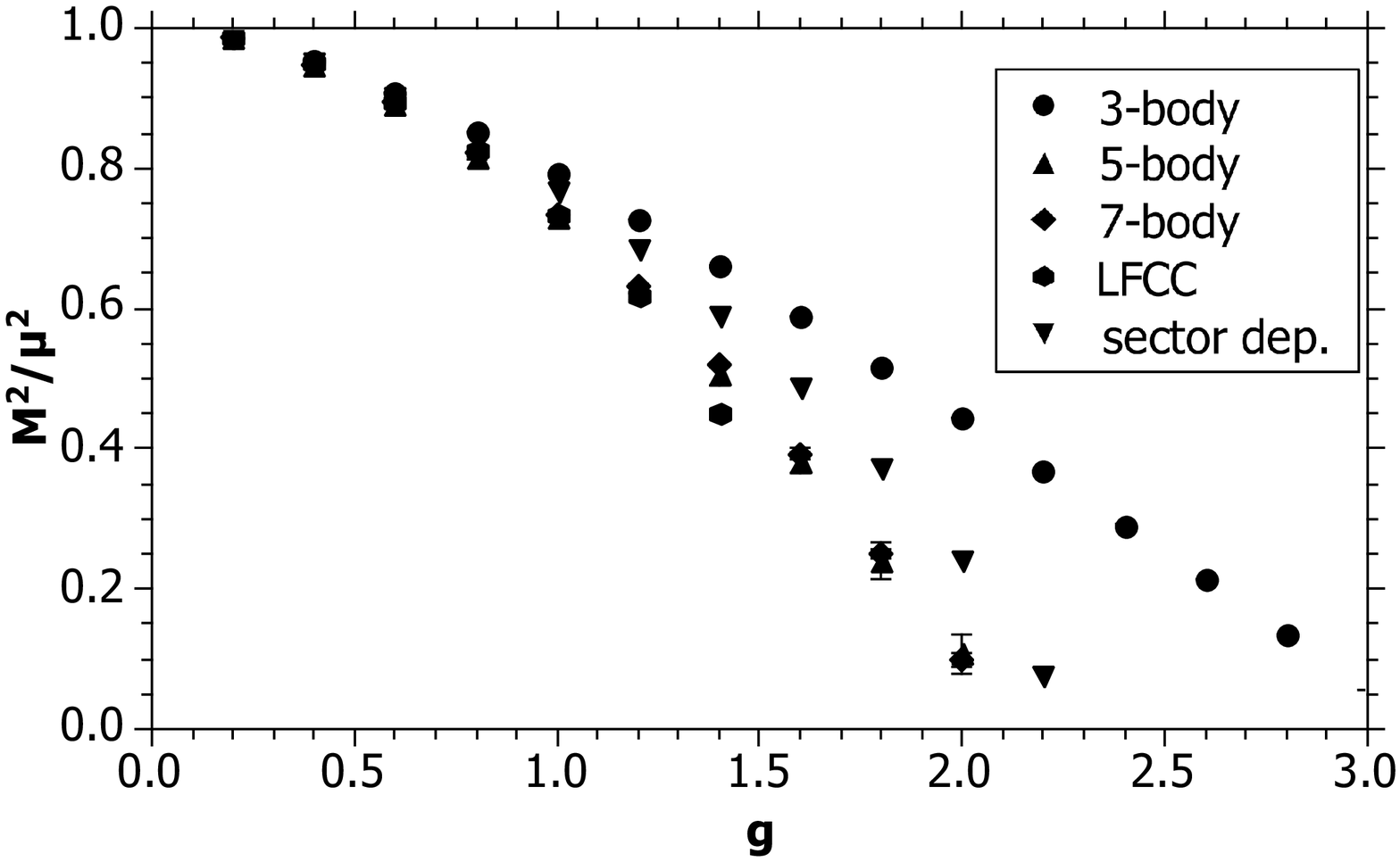} &
\includegraphics[width=7cm]{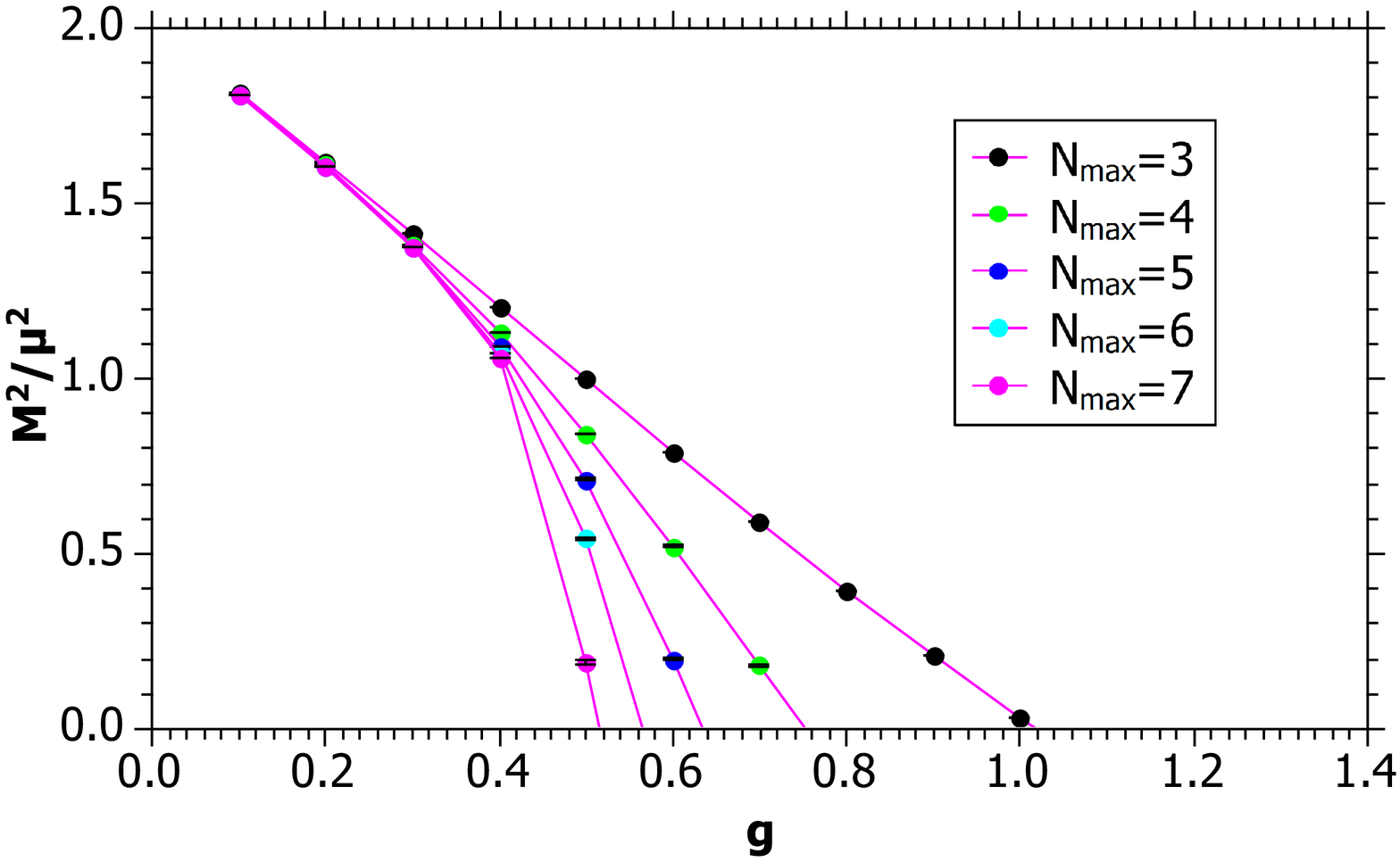} \\
(a) & (b) 
\end{tabular}
\caption{Mass squared vs coupling strength for the (a) symmetric phase
and (b) broken phase.  The different Fock-space
truncations in (a) are the three-body (triangles), five-body
(squares), and seven-body (diamonds) Fock sectors.  Results
for the light-front coupled-cluster method~\protect\cite{LFCCphi4}
(circles) are also included.
In (b), each set of points corresponds
to a different Fock-space truncation to $\Nmax$ constituents.
the different truncations are the four-body (triangles),
six-body (squares), and eight-body (diamonds) Fock sectors.
Error bars are determined by the fits to extrapolation in
the polynomial basis size.}
\label{fig:M2vsg}
\end{figure}

Figure~\ref{fig:M2vsg} shows the results for the 
mass squared of the lowest eigenstate as a function
of the dimensionless coupling $g$ and for both the
symmetric and broken phases.
In the symmetric phase, we find a critical coupling of~\cite{phi4sympolys}
$g=2.1\pm0.05$.  In the broken phase, the critical coupling values extrapolate 
to $g=0.2\pm0.02$.

\begin{figure}
\centering
\includegraphics[width=10cm]{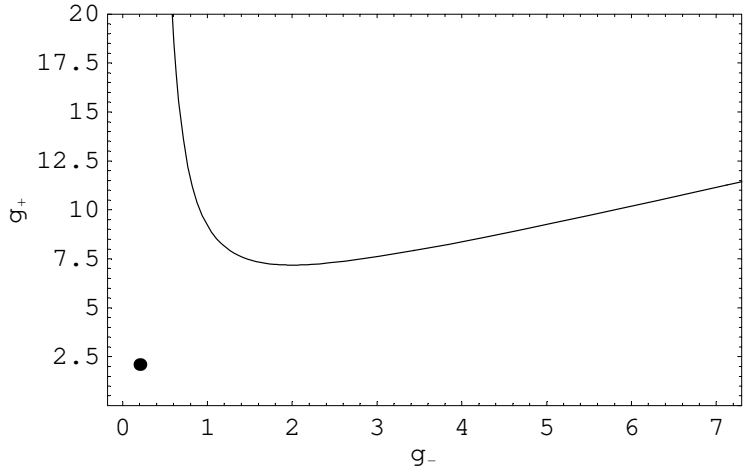}
\caption{The duality in couplings between the symmetric and broken phases.
The solid line is the semi-classical duality of Chang~\protect\cite{Chang2}.
The point corresponds to our numerical results.}
\label{fig:dual}
\end{figure}

These critical coupling values can be compared to Chang's duality~\cite{Chang1,Chang2,Kim,HariVary},
which is obtained from the connection between normal orderings with respect to different
masses~\cite{normal}:
\bea
N_+[\phi^2]&=&N_-[\phi^2]+\frac{1}{4\pi}\ln\frac{\mu_+^2}{\mu_-^2},\\
N_+[\phi^4]&=&N_-[\phi^4]+6\frac{1}{4\pi}\ln\frac{\mu_+^2}{\mu_-^2}N_-[\phi^2]
             +3\left(\frac{1}{4\pi}\ln\frac{\mu_+^2}{\mu_-^2}\right)^2.
\eea
The Hamiltonian density is then written as
\be
{\cal H}^-=\left(\frac12\mu_+^2+\frac{\lambda}{4}\frac{1}{4\pi}\ln\frac{\mu_+^2}{\mu_-^2}\right)N_-[\phi^2]
       +\frac{\lambda}{4!}N_-[\phi^4] 
 +\frac{1}{4\pi}\ln\frac{\mu_+^2}{\mu_-^2}
           \left(2\mu_+^2+\frac{\lambda}{8}\frac{1}{4\pi}\ln\frac{\mu_+^2}{\mu_-^2}\right).
\ee
This is equivalent to ${\cal H}^-$ with negative mass squared if
$\frac12\mu_+^2+\frac{\lambda}{4}\frac{1}{4\pi}\ln\frac{\mu_+^2}{\mu_-^2}=-\frac12\mu_-^2$.
For the dimensionless couplings $g_\pm\equiv \lambda/4\pi\mu_\pm^2$ this becomes
$\frac{1}{g_+}-\frac12\ln g_+=-\frac12\ln (g_-)-\frac{1}{g_-}$.
The comparison is illustrated in Fig.~\ref{fig:dual}

The relative probabilities of the higher Fock sectors are readily computed
from the Fock-state wave functions obtained in the eigenvector of the
Hamiltonian matrix.  A plot can be found in \cite{ChabyshevaLC16}.
There is no indication of critical behavior at $g=2.1$, where one
would expect that the higher Fock sectors would dominate.  A 
natural assumption for why this might be happening is that the
numerical calculation used the same constituent mass $\mu$ in every
Fock sector, making the invariant mass of the $N$-constituent
Fock sector of order $\sum_i^N\frac{\mu^2}{1/N}=N^2N\mu^2$.  The higher 
sectors are then suppressed by this large invariant mass.  

\begin{figure}
\centering
\begin{tabular}{cc}
\includegraphics[width=7cm]{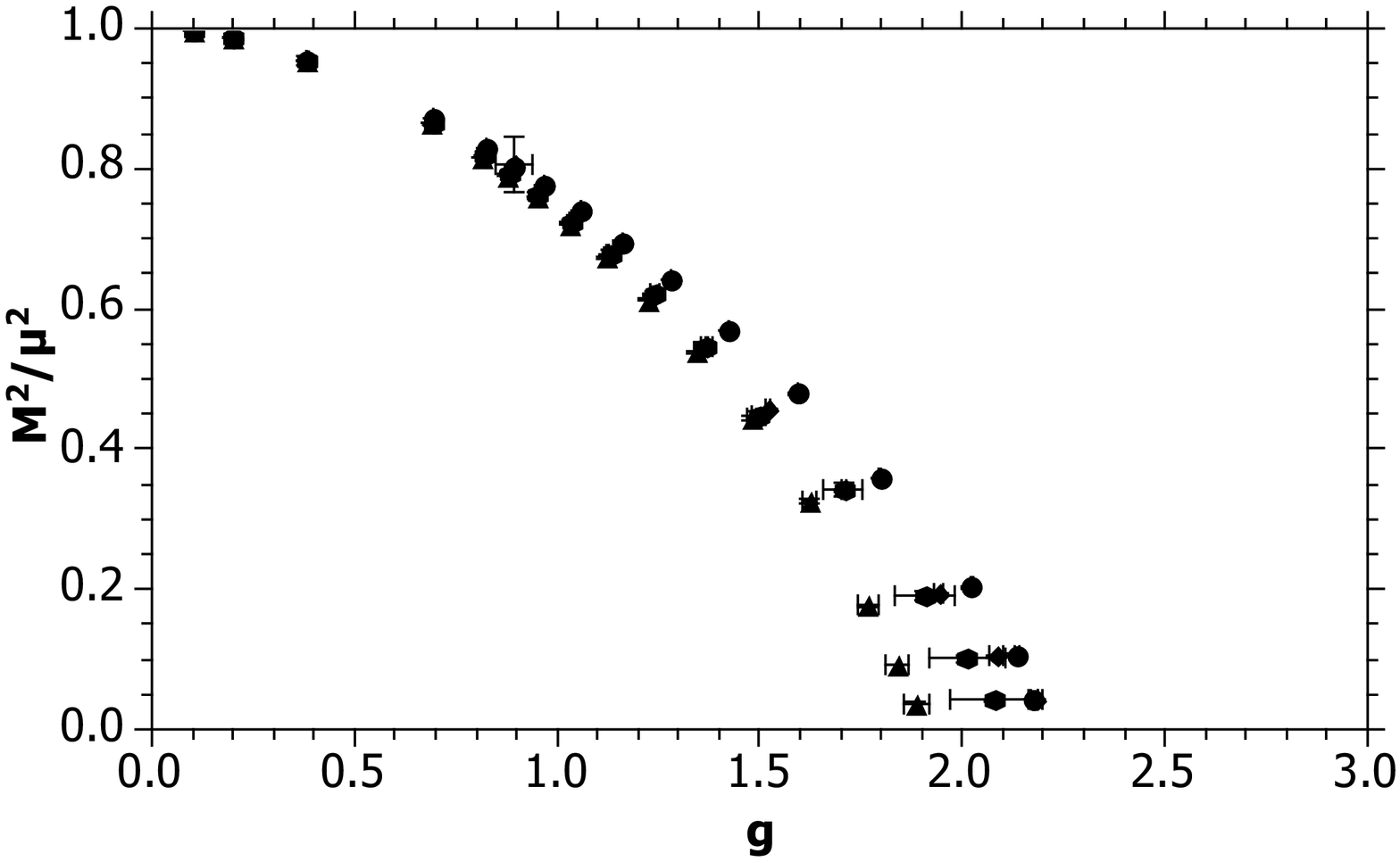} &
\includegraphics[width=7cm]{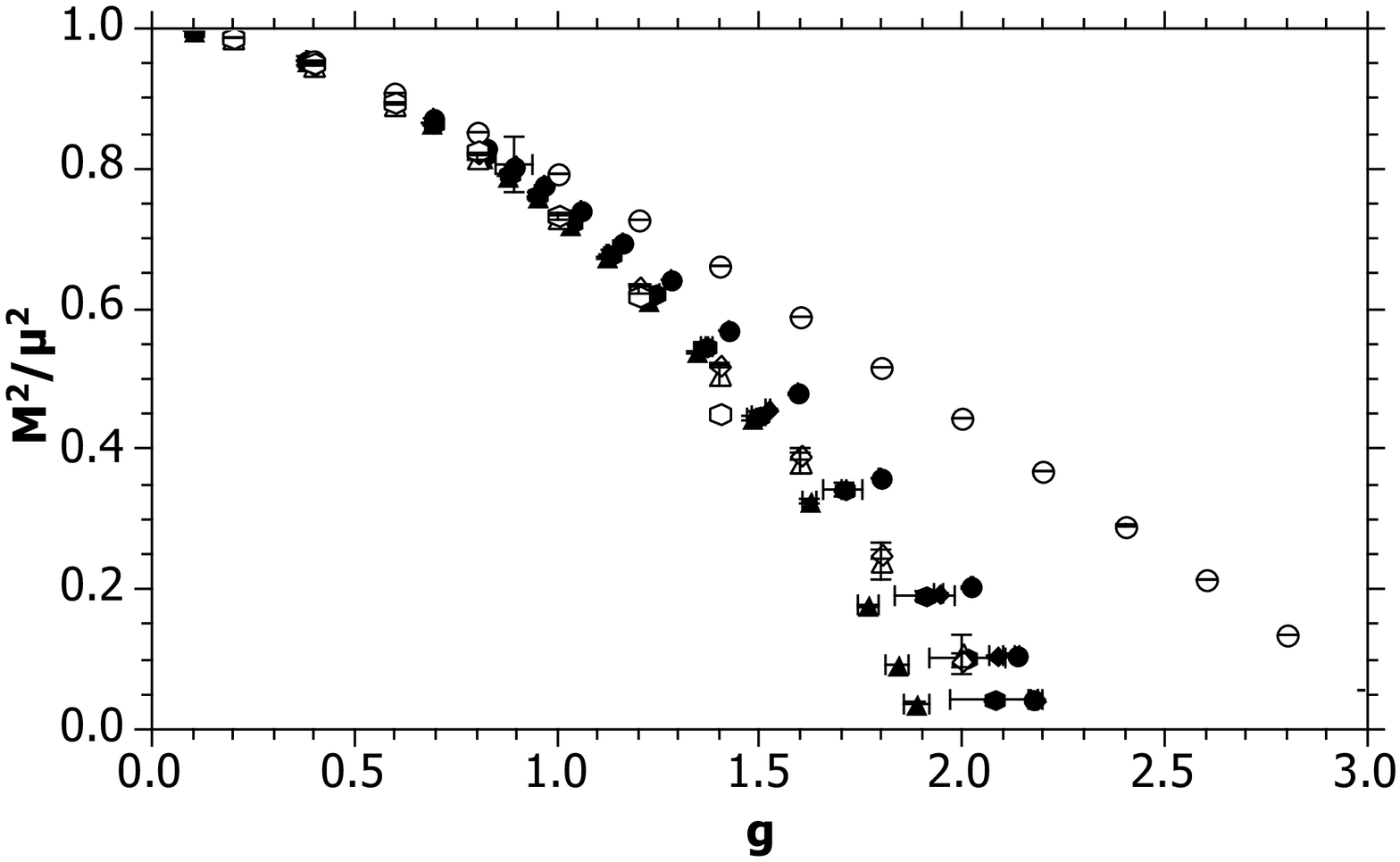} \\
(a) & (b) 
\end{tabular}
\caption{Mass squared for the lowest eigenstate as a function of the
dimensionless coupling $g$ with (a) a sector-dependent constituent mass
and (b) with both sector-dependent (filled symbols) and independent (open symbols).
For (a), the Fock-space truncations are to a number of constituents
$\Nmax=3$ (circles), 5 (triangles), 7 (diamonds), and 9 (hex); the
error bars estimate the range of fits for the $\mu_1$ extrapolations
used to obtain $g$ and $M$.  In (b), the sector-independent,
five and seven-constituent results are nearly identical
with the nine-constituent sector-dependent results. Plot (b)
also includes, as open hexagons, the results from a light-front
coupled-cluster calculation~\protect\cite{LFCCphi4}.
}
\label{fig:extrap}
\end{figure}

This can be avoided by the use of a sector-dependent constituent
mass~\cite{SecDep1,SecDep2,SecDep3,SecDep4,SecDep5,SecDep6}; 
our approach is described in \cite{ChabyshevaLC16}.
The results for the lowest odd eigenstate are shown in Fig.~\ref{fig:extrap}.
They are consistent with the sector-independent approach, and the estimate 
of the critical coupling remains unchanged at a value of 2.1.
As can be seen in \cite{ChabyshevaLC16}, the relative probabilities
also remain nearly the same.

\section{Summary}  \label{sec:summary}

These calculations have revealed two inconsistencies
in the light-front approach to $\phi^4$ theory.   One is the 
absence of a vacuum expectation value (VEV) above the critical 
coupling, where the $\phi\rightarrow-\phi$ symmetry is to
be broken, and the presumably related behavior in
the explicitly broken phase, where the vacuum
expectation value remains nonzero above the critical
coupling.  A Gaussian effective 
potential analysis~\cite{GEP1,GEP2} can determine a VEV that does
flip between zero and non-zero, but it does so discontinuously,
which is inconsistent with the known second-order nature
of the transition.

Another inconsistency is the smooth behavior of the computed Fock sector
probabilities as the coupling is increased through the critical value,
as presented in more detail in \cite{ChabyshevaLC16}. 
The relative probabilities for the sector-dependent and independent
calculations are essentially the same in the three-body Fock sector.
This indicates full convergence with respect to the Fock-space truncation.
In Fock sectors with five and seven constituents, the relative 
probability for the sector-dependent case rises above the probability
in the standard case as the critical coupling is approached.
This greater probability is expected; however, the full expectation 
was that these probabilities would increase much more rapidly.
The hypothesis, that a sector-dependent mass would reveal the critical 
behavior, must be incorrect.  It seems likely that a coherent-state
approach is needed, something that the light-front coupled cluster
method can bring~\cite{LFCC,LFCCphi4}.

\acknowledgments
This work was done in collaboration with S.S. Chabysheva and was
supported in part by the Minnesota Supercomputing Institute
of the University of Minnesota with allocations of computing
resources.

\end{document}